\documentclass[preprint,prd,nofootinbib,tightenlines]{revtex4}
\usepackage{epsfig}
\usepackage{graphicx}
\usepackage{dcolumn}
\usepackage{bm}
\usepackage{color}

\begin{document}
\baselineskip=15pt \parskip=5pt

\title{What do we learn from the $\rho-\pi$ puzzle}

\author{Xue-Qian Li}
\email{lixq@nankai.edu.cn}

\affiliation{Department of Physics, Nankai University, Tianjin,
300071}

\date{\today}

\begin{abstract}
The experimental observation indicates that the branching ratio of
$\psi'\to \rho\pi$ is very small while the $\rho-\pi$ channel is a
main one in $J/\psi$ decays. To understand the puzzle, various
interpretations have been proposed. Meanwhile according to the
hadronic helicity selection rule the decay mode $J/\psi\to\rho\pi$
should be suppressed, but definitely, numerical computation is
needed to determine how it is suppressed. We calculate the branching
ratios corresponding to subprocesses $J/\psi\to ggg\to\rho\pi$ and
$\to ggg\to\pi\pi$ in the framework of QCD. The results show that
the branching ratios are proportional to $({m_u+m_d\over
M_{J/\psi}})^2$ for $\rho\pi$ mode and $({m_u-m_d\over
M_{J/\psi}})^2$ for $\pi\pi$ mode which is an isospin-violation
channel. If only the OZI process is considered, the theoretical
prediction on the ratio of $J/\psi\to \rho\pi$ is smaller than data,
but not too drastically small. Meanwhile, a possible interpretation
for the $\rho\pi$ puzzle is proposed that the suppression is due to
interference between OZI and electromagnetic (EM) contributions.
Thus based on this observation, we suggest that if the amplitudes of
the strong OZI process via an s-channel three-gluon intermediate
state and electromagnetic one via an s-channel virtual photon
intermediate state have the same order of magnitude, and
constructively interfere for $J/\psi\to\rho\pi$, but destructively
interfere for $\psi'\to\rho\pi$, thus simultaneously the $\rho\pi$
puzzle disappears and  the sizable width of $J/\psi\to\rho\i$ is
understandable. However, so far, we cannot derive the phase
difference from an underlying principle of QCD yet. Alternative
interpretations are also discussed in the text.

\end{abstract}

\maketitle

\section{Introduction}
The $\rho-\pi$ puzzle has been standing for years. The puzzle is
phrased that $J/\psi\to\rho\pi$ is a main channel in $J/\psi$
decays, while the branching ratio of $\psi'\to\rho\pi$ is very
small. It seems to contradict to the general understanding of the
charmonia physics.

In the regular theoretical framework, there should be a relation
$$R={BR(\psi'\rightarrow ggg)\over BR(J/\psi\rightarrow
ggg)}={\Gamma(\psi'\rightarrow e^+e^-)\over \Gamma(J/\psi\rightarrow
e^+e^-)}\cdot {\Gamma_t(J/\psi)\over \Gamma_t(\psi')},$$ where
$\Gamma_t$ is the total width.   This estimate on the ratio
originates from the fact that if both $J/\psi$ and $\psi'$ are
$c-\bar c$ bound states, as commonly conjectured, in the hadronic
decays, $c$ and $\bar c$ annihilate into three gluons which then
convert into hadrons, whereas in the leptonic decays, $c$ and $\bar
c$ annihilate into a virtual photon which turns into a lepton-pair.
In this picture, the amplitudes of the hadronic decays which are
supposed to occur via a three-gluon intermediate state, and the
leptonic decay which occurs via a virtual photon intermediate state
are proportional to the wavefunction at origin $\psi(0)$. If
everything worked well, the ratio should be close to 12$\sim 14$\%,
which is called as the 14\% rule (now, it is sometimes called as
12\% rule, anyhow it is a sizable number.). However the data tell us
that this ratio is much smaller than this value\cite{Databook}.

Some theoretical interpretations have been proposed. Rosner et al.
\cite{Rosner} suggested that the observed $\psi'$ may be not a pure
2S state which is the first radial excited state of the $c\bar c$
system, but a mixture of 2S and 1D states. The amplitudes are
instead
$$\langle\rho\pi|\psi'\rangle=\langle\rho\pi|2^3S_1\rangle\cos\phi-
\langle\rho\pi|1^3D_1\rangle\sin\phi\sim 0,$$
$$\langle\rho\pi|\psi"\rangle=\langle\rho\pi|2^3S_1\rangle\sin\phi+
\langle\rho\pi|1^3D_1\rangle\cos\phi\sim
\langle\rho\pi|2^3S_1\rangle/\sin\phi,$$ where the mixing angle
$\phi$ is fixed as $-27^{\circ}$ or $12^{\circ}$ by fitting data. By
a destructive interference between the contributions of the two
components to the amplitude of $\psi'\rightarrow \rho\pi$, the
smallness is explained. Suzuki \cite{Suzuki} alternatively suggested
that the relative phase between the one-photon and gluonic decay
amplitudes in $\psi'$ decay may result in the small branching ratio.
The final state interactions may also give a reasonable explanation
\cite{Zou-11}. The first proposal can be tested in the decays of
$\psi"\rightarrow \rho\pi$ which has not been well measured yet. In
Ref. \cite{Suzuki}, the author suggested that the one-photon
amplitude is sizable and it can be tested in some other modes, for
example $\psi'\rightarrow \pi\pi$ if the process is dominated by the
electromagnetic interaction. The hadronic excess can also receive
tests in the decays of other higher excited states of $\psi-$ family
and even $\Upsilon-$ family. We will come to this important point at
the discussion section below.

Another puzzle is raised for $J/\psi\to\rho\pi$ decay, it is
generally believed that $J/\psi$ hadronic decays are dominated by
strong interaction, namely the hadronic decay processes occur via an
s-channel three-gluon intermediate state which eventually hadronize
into hadrons. Those are the OZI-forbidden processes. More seriously,
as Brodsky et al. indicated, there is a so-called helicity selection
rule\cite{Brodsky}, which forbids the process of $J/\psi\to \rho\pi$
as long as the masses of light quarks can be neglected. The helicity
section rule, in fact, would greatly suppress the direct decay of
$J/\psi\to \rho\pi$, even though the masses of u and d quarks are
not set to zero. However, the data show that it is one of the main
channels of $J/\psi$ decays and it demands an explanation along with
the $\rho\pi$ puzzle for $\psi'$.

The final state interaction may play an important role as it does
for $D$ and $B$ decays, in decays of $\psi$ charmomia as suggested
in our earlier work\cite{Liu}.  The final state interaction process
is induced by strong interaction at lower energy region, thus it is
governed by the non-perturbative QCD which is not fully understood
in the present theoretical framework yet. People need to invoke some
phenomenological models to carry out the calculations. In our work
\cite{Liu}, in terms of a simple model, we simultaneously consider
the FSI and the direct decay of $J/\psi$ into a vector meson plus a
pseudoscalar meson and conclude that both of them contribute to the
widths and their interference should be destructive to explain data.
The difficulties are how to properly evaluate such effects. A
detailed discussion on estimation of final state interaction was
presented in our previous works, here we only cite our results to
discuss the puzzle.

\section{Estimate of the contribution from the OZI process}

The first step is properly evaluating the contribution from the OZI
process. In a straightforward calculation based on the SM, we
estimate the decay width of the OZI forbidden process $J/\psi\to
ggg\to\rho\pi$ \cite{Li-Tong}, and find that the width is indeed
proportional to $(m_q/m_{J/\psi})^2$ which stands as the hadronic
helicity suppression factor. Numerically the branching ratio of
$J/\psi\to\rho\pi$ should be smaller than 0.1\%. The same situation
appears for $\psi'\to\rho\pi$. To testify the calculation, we
recalculate the subprocess $J/\psi\to ggg\to \pi\pi$, which is an
isospin violating reaction and usually is supposed to be dominated
by the subprocess $J/\psi\to\gamma^*\to\pi\pi$ because the EM
interaction violates isospin as well known. Our result indicates
that in the OZI forbidden subprocess the transition amplitude is
proportional to $(m_u-m_d)/m_{J/\psi}$, i.e. the mass difference
results in the isospin violation instead.  All these consequences
are consistent with our physics picture and qualitatively
reasonable. Therefore we can trust our calculations for the process
$J/\psi\to ggg\to\rho\pi$. Our numerical results are listed in Table
\ref{aa}.
\begin{table}[h]
\caption{Decay widths ($\Gamma$) of $J/\psi\to \pi^+ \rho^- + \pi^-
\rho^+$ based on the three distribution functions, $\phi_1$,
$\phi_2$ and $\phi_3$, respectively \cite{Li-Tong}.  \label{aa}}
\begin{center}
\begin{tabular}{c|c|c|c|c|c}
  \hline\hline
  $m_u$(MeV) & $m_d$(MeV) & $\Gamma(\phi_1)$(MeV) & $\Gamma(\phi_2)$(MeV)
  & $\Gamma(\phi_3)$(MeV) & exp(MeV) \\
  \hline\hline
  2 & 2 & $1.04\times 10^{-4}$ & $7.21\times 10^{-5}$ & $5.11\times 10^{-4}$ &  \\
  3 & 3 & $2.36\times 10^{-4}$ & $1.6\times 10^{-4}$ & $1.17\times 10^{-3}$ &  \\
  4 & 4 & $4.12\times 10^{-4}$ & $2.9\times 10^{-4}$ & $2.08\times 10^{-3}$
  &  $(1.06\pm 0.08)\times 10^{-3}$\\
  5 & 5 & $6.69\times 10^{-4}$ & $4.54\times 10^{-4}$ & $3.38\times 10^{-3}$ &  \\
  6 & 6 & $9.75\times 10^{-4}$ & $6.68\times 10^{-4}$ & $4.88\times 10^{-3}$ &  \\
  \hline\hline
\end{tabular}
\end{center}
\end{table}

Our numerical result for $J/\psi\to ggg\to \pi\pi$ is of the same
order as the data. On other aspect, the EM sub-process is also
responsible for the isospin violating decay mode, i.e. the
contribution of $J/\psi\to\gamma^*\to\pi\pi$ should also be of the
order of the data. This observation implies that the contribution of
the OZI and EM sub-processes should have the same order of
magnitude, even though the details of the theoretical estimate are
somehow model-dependent. This observation would support our proposal
for explaining the $\rho\pi$ puzzle and dismissing the suppression
from the helicity conservation (see below discussion).

This is not surprising because the relative ratio between the OZI
and EM contribution is roughly proportional to
$$ {{\alpha_s^3\over \pi}\over\alpha \kappa}\sim 1.1, $$
where $1/\pi$ comes from the loop integration, the EM coupling
$\alpha\sim 1/137,$ and at the $J/\psi$ energy scale, $\alpha_s$
could be around 0.3 $\kappa$ represents a numerical factor of order
O(1) for a quark-pair production from vacuum. Definitely, this ratio
depends on the value of the running constant $\alpha_s$. Since in
the energy region of charm mass the non-perturbative QCD effects
begin to play roles, the value of $\alpha_s$ should have an
uncertainty, that is what we refer above as the details in
theoretical calculations.

\section{Discussion}

As indicated in \cite{Brodsky}, to understand the sizable rate of
$J/\psi\to\rho\pi$ which should be suppressed by the hadronic
selection rule, the structure of $J/\psi$ may be not a pure $c\bar
c$ charmonium, but consists of other components, such as hybrids
$c\bar cg$, $c\bar cq\bar q$ and etc.

For another aspect, to understand the smallness of the ratio $R$, it
is suggested that there is a possible mechanism which would suppress
$\psi'\to\rho\pi$, but  does not much influence on $J/\psi$, or a
completely different alternative picture is that there may be
something obscure in $J/\psi$ structure while the smallness of
$\psi'\to\rho\pi$ is normal. Our above numerical results show that
even though the hadronic selection rule works in the cases of
$J/\psi\to \rho\pi $ and $\psi'\rightarrow \rho\pi$, the suppression
is not too serious and the theoretical prediction is only smaller
than the data by less than one order.

A theory should be raised to compromise the both anomalies.

The proposed mixing structures of $\psi'$ and $\psi"$ may still be a
possibility for the $\rho\pi$ puzzle, but it needs further
experimental test. However, it cannot explain the enhancement of
$J/\psi\to\rho\pi$ which should be suppressed by the hadronic
helicity conservation and more complicated mechanisms may be needed.

The final state interaction (FSI) might cause a suppression which
induces the small R-value, and moreover it also enhances the
branching ratio of $J/\psi\to\rho\pi$\cite{Liu}, but because the
theoretical estimate depends on several phenomenological input
parameters, it is not very accurate. Thus one cannot be fully
convinced that FSI is the final answer yet.

Coming to the anomaly caused by the helicity selection rule, if the
$J/\psi$ is not a pure $c-\bar c$ bound state, this would be a great
challenge to our understanding because the $c\bar c$ structure of
$J/\psi$ has been recognized almost from very beginning of its
discovery. If it is really true, all the previous works based on the
potential models where many parameters are fixed by fitting data
should be re-considered.

To alleviate the constraint from the helicity conservation, there
may be some other mechanisms which were not taken into account, or
may exist contributions from new physics beyond the standard model
(SM). However, the later seems not very promising because the
concerned energy range is rather low and SM works perfectly well to
explain the data for all processes and almost no room remains for
new physics.

Pretty interesting, there is also an alternative opinion towards the
subject, Suzuki \cite{Suzuki}, Zhao \cite{Zhao-Qiang} deny the small
R-value as a "puzzle", because they consider that the
electromagnetic interaction may play an important role in $\psi'$
decays where $c\bar c$ annihilate into a virtual photon which later
fragment into hadrons. In the picture, it is supposed that a
destructive interference between the contribution of three-gluon and
single-photon processes would suppress $\psi'\to\rho\pi$.

Following this idea,  we further propose an alternative possibility,
namely interferences between the OZI and EM contributions which have
the same order of magnitude, result in not only small R, but also
the measured rate of $J/\psi\to\rho\pi$.

As indicated above by our calculation, the process $J/\psi\to\pi\pi$
is suppressed by isospin violation, two main contributions to the
process, i.e. the OZI and EM contributions should have the same
order of magnitude. And by our aforementioned simple estimate of
order of magnitude, one can be convinced that the OZI and EM
contributions to $J/\psi\to\rho\pi$ and $\psi'\to\rho\pi$ may also
have the same order of magnitude. Thus even though the contribution
from the OZI process to $J/\psi\to ggg\to\rho\pi$ is smaller than
the data, a constructive interference with the EM contribution may
double the branching ratio and the final result is close to the
data. Meanwhile, a destructive interference greatly diminishes the
branching ratio of $\psi'\to\rho\pi$. Since the two types of
contributions are close to each other in magnitude, the destruction
may explain data.

There have been some other theoretical explanations besides that we
discussed above, in the work by Mo, Yuan and Wang \cite{X-Mo}, the
authors described the recent status of theoretical research as well
as the experimental measurements on the interesting subject.

As a brief summary, we can list several interpretations for the $
rho\pi$ puzzle and hadronic helicity constraint in the framework of
the standard model.

First one is that $\psi'$ is a mixing of 2s and 1d states and
meanwhile $J/\psi$ may be a high Fock state, such as a hybrid or
contains a sizable component of hybrid.

Secondly, the final state interaction may explain the sizable
branching ratio of $J/\psi\to\rho\pi$ and suppression of
$\psi'\to\rho\pi$.

Third one is that a constructive interference between the OZI and EM
contributions enhances the branching ratio of $J/\psi\to\rho\pi$ and
their destructive interference remarkably diminishes the branching
ratio of $\psi'\to\rho\pi$. This interpretation may be more
reasonable, but it still suffers from the argument that a fine
tuning is required to result in a very small (by orders) branching
ratio of $\psi'\to\rho\pi$ and so far we do not have a convincing
calculation to show why it is constructive for $J/\psi\to\rho\pi$,
but destructive for $\psi'\to\rho\pi$.

Of course, the three mechanisms may all exist and contribute to the
double "puzzles" altogether. We need to design new experiments to
testify all the possibilities. Thanks to the newly operating BEPC II
and BES III, the high luminosity and detection precision provide
such opportunities to carry out very accurate measurements by which
we may draw more definite conclusion about the above interpretations
or suggest new ones.\\

Acknowledgements:

We are greatly benefited from discussion with Dr. Q. Zhao, in fact
we have learnt the idea about the interference between the OZI and
EM processes from the discussion with him. This work is supported by
the National Natural Science Foundation of China and the Special
Grant of the Education Ministry of China. This paper will appear at
the proceeding of the Conference on advanced topics in the
interdisciplinary fields of particle physics, nuclear physics and
cosmology, Tenchong, Yunan province, China, held from Aug.1 to
Aug.5.

\end{document}